\documentclass[iop]{emulateapj}
\usepackage[latin1]{inputenc}
\usepackage[english]{babel}
\usepackage{amsmath}
\usepackage{amssymb}
\usepackage{latexsym}
\usepackage{epsfig}
\usepackage{subfigure}
\usepackage{natbib}
\usepackage{setspace}
\usepackage{graphicx}
\usepackage{longtable}
\shorttitle{Binary astrometric microlensing with Gaia}
\shortauthors{Sajadian}
\begin{document}

\title{Binary astrometric microlensing with Gaia}
\author{Sedighe Sajadian \altaffilmark{1,2}}
\altaffiltext{1}{School of Astronomy, Institute for Research in
Fundamental Sciences (IPM), P.O. Box 19395-5531, Tehran,
Iran}\email{sajadian@ipm.ir} \altaffiltext{2}{Department of Physics,
Sharif University of Technology, P.O. Box 11155-9161, Tehran, Iran}

\begin{abstract}
We investigate whether Gaia can specify the binary fractions of
massive stellar populations in the Galactic disk through astrometric
microlensing. Furthermore, we study if some information about their
mass distributions can be inferred via this method. In this regard,
we simulate the binary astrometric microlensing events due to
massive stellar populations according to the Gaia observing strategy
by considering (i) stellar-mass black holes, (ii) neutron stars,
(iii) white dwarfs and (iv) main-sequence stars as microlenses. The
Gaia efficiency for detecting the binary signatures in binary
astrometric microlensing events is $\sim 10-20$ per cent. By
calculating the optical depth due to the mentioned stellar
populations, the number of the binary astrometric microlensing
events being observed with Gaia with detectable binary signatures,
for the binary fraction about $0.1$, is estimated as $6$, $11$, $77$
and $1316$ respectively. Consequently, Gaia can potentially specify
the binary fractions of these massive stellar populations. However,
the binary fraction of black holes measured with this method has the
large uncertainty owing to a low number of the estimated events.
Knowing the binary fractions in massive stellar populations helps
for studying the gravitational waves. Moreover, we investigate the
number of massive microlenses which Gaia specifies their masses
through astrometric microlensing of single lenses toward the
Galactic bulge. The resulted efficiencies of measuring the mass of
mentioned populations are $9.8$, $2.9$, $1.2$ and $0.8$ per cent
respectively. The number of their astrometric microlensing events
being observed in the Gaia era in which the lens mass can be
inferred with the relative error less than $0.5$ toward the Galactic
bulge is estimated as $45$, $34$, $76$ and $786$ respectively.
Hence, Gaia potentially gives us some
information about the mass distribution of these massive stellar populations.\\
\end{abstract}

\keywords{Gravitational lensing: micro - Astrometry- Stars:
binaries: general- methods: numerical}

\section{Introduction}
% ---------------- introducing Gaia ---------------------------------------------
Gaia is a space observatory of the European Space Agency (ESA),
performing high precision astrometry, multi-colour, multi-epoch
photometry and spectroscopy on astronomical objects brighter than
$20$ G magnitudes. This satellite, surveying the throughout sky,
will observe each object $72$ times averagely during five years of
its lifetime \citep{Eyer2013}. The astrometric precision of Gaia
depends strongly on the source magnitude and will be $30\mu$as to
$600\mu$as for a single astrometric measurement of stars with
magnitude $13$ to $19$ \cite{Varadi2009}. Different astronomical
phenomena could be observed during the Gaia mission, e.g. supernovae
\cite{Belokurov2003}, novae, dwarf novae, eruptive stars, variable
stars \cite{Varadi2009,Eyer2013}, astrometric microlensing events
\cite{Belokurov2002,Proft2011}, etc. Here, we study some features of
detecting the last one with Gaia.

% ---------------- introducing astrometric microlensing ------------------------
If the light of a background star passes through the gravitation
field of a collinear foreground object, according to the Einstein
general relativity, the light beam bends toward the center of
gravity \cite{Einstein36}. Consequently, several distorted,
(un)magnified images are produced whose angular separations are too
small to be resolved even by modern telescopes. Instead, the
combined and magnified light of images is received by observer,
so-called gravitational microlensing. One of its feature is the
displacement between the light centroid of images and the source
position while the source star is passing the gravitational field of
lens, i.e. astrometric trajectory
\cite{Walker,Miyamoto,Hog,Jeong}. During a microlensing event by
measuring the astrometric lensing and parallax effect, the mass of
the deflector can be inferred \cite{Paczynski97,Miralda96}. In
microlensing events, the astrometric cross-section is much larger
than the photometric one \cite{Paczynski96}. Hence, the optical
depth of astrometric microlensing events is larger than the
photometric one \cite{DominikSahu}.

% ---------------- previous works -----------------------------------------------
Detecting the astrometric microlensing with the Gaia satellite was
first discussed by Dominik \& Sahu (2000). Then, Belokurov and Evans
(2002) studied this subject by more details.% and estimated that in
%the Gaia era about $25000$ astrometric microlensing events due to
%main-sequence stars with the significant variations of the centroid
%shift in the source position will be observed.
The astrometric microlensing due to massive objects in solar orbit
beyond the Kuiper Belt was investigated by Gaudi and Bloom (2005).
Also, Proft et al. (2011) investigated several catalogs to find
several nearby stars with large proper motions which are potential
candidates for the astrometric microlensing detection in the Gaia
era.

%-------------- Our aim ----------------------------------------------------------
Here, we investigate quantitatively the Gaia efficiency for
specifying the \emph{binarity fractions of microlenses} in binary
astrometric microlensing due to massive stellar populations, i.e.
stellar-mass black holes, neutron stars and white dwarfs in addition
to main-sequence stars. In order to obtain the Gaia efficiency, we
preform a Monte Carlo simulation. Then, we evaluate the optical
depth and the number of binary events being observed with Gaia with
detectable binary signatures, considering different binary fractions
for massive stellar populations. Indicating the binary fractions in
massive stellar populations is useful for studying the gravitational
waves and estimating their amounts, see e.g. Riles (2013), Cutler
and Thorne (2002). We find that Gaia can potentially point out the
binary fractions of massive stellar populations. Furthermore, we
estimate the Gaia efficiency for measuring the mass of
\emph{massive} stellar populations in the Galactic disk through
astrometric microlensing of a single lens to indicate if Gaia can
give any information about their mass distributions. Indeed,
astrometric microlensing events due to more massive microlenses have
longer Einstein crossing times and larger angular Einstein radii.
Therefore, the Gaia efficiency for measuring the lens mass in these
events is high.

This paper is organized as follows: In section (\ref{one}) the
basics of astrometric microlensing are explained. In the next
section, we simulate binary astrometric microlensing events being
observed with Gaia to investigate how much information Gaia gives us
about the binary stellar populations in the Galactic disk. In
section (\ref{three}) we study if Gaia can determine the mass
distributions of massive stellar populations through detecting
astrometric microlensing of a single lens. We conclude in the
last section.\\

%%%%%%%%%%%%%%%%%%%%%%%%%%%%%%%%%%%%%%%%%%%%%%%%%%%%%%%%%%%%%%%%%%%%%%%%%%%%%%%%%%%%%%%%%%%%%%%%%%%%%%%%%5
\section{Basics of Astrometric microlensing}\label{one}
In microlensing events, the light centroid vector of source star
images does not coincide with the source position. This astrometric
shift in source star position changes with time as a microlensing
event progresses. For a point-mass lens, the centroid shift vector
of source star images is given by \cite{Walker,Miyamoto,Hog}:
\begin{eqnarray}\label{astro1}
{\bf \delta}{\bf \theta}_{c}=\frac{\mu_{1}{\bf \theta}_{1}+
\mu_{2}{\bf \theta}_{2}}{\mu_{1}+\mu_{2}}-{\bf u}\theta_{E}
=\frac{\theta_{E}}{u^{2}+2}{\bf u},
\end{eqnarray}
where ${\bf \theta}_{i}$ and $\mu_{i}$ are the position and
magnification factor of $i$th image, ${\bf u}=p\hat{x}+u_{0}\hat{y}$
is the vector of the projected angular position of the source star
with respect to the lens normalized by the angular Einstein radius
of the lens $\theta_{E}$ in which $p=(t-t_{0})/t_{E}$ , $t_{0}$ is
the time of the closest approach, $t_{E}$ is the Einstein crossing
time, $\hat{x}$ and $\hat{y}$ are the unit vectors in the directions
parallel with and normal to the direction of the lens-source
transverse motion. The angular Einstein radius of lens is given by:
\begin{eqnarray}\label{Eins}
\theta_{E}=\sqrt{\kappa ~M_{l} ~\pi_{rel}} = 300 \mu
as\sqrt{\frac{M_{l}}{0.3 M_{\odot}}}
\sqrt{\frac{\pi_{rel}(mas)}{0.036}},
\end{eqnarray}
where $M_{l}$ is the lens mass, $\kappa=4G/(c^{2} A.U.)$ and
$\pi_{rel}=A.U.(\frac{1}{D_{l}}-\frac{1}{D_{s}})$ where $D_{l}$ and
$D_{s}$ are the lens and source distances from the observer.
Usually, a microlensing parallax is defined as $\pi_{E}=\pi_{rel}/
\theta_{E}$ which can be measured from the photometric event.
According to equation (\ref{Eins}), by measuring the relative
parallax $\pi_{rel}$ and the angular Einstein radius from
astrometric observations the lens mass can be inferred. Nearby
lenses have considerable and measurable relative parallaxes and the
angular Einstein radii which are suitable candidates for astrometric
measurements with Gaia. In that case, if the lens mass is high, the
error bars of the lenses mass strongly decrease due to their large
angular Einstein radii.

In astrometric microlensing, the threshold amount of impact
parameter i.e. $u_{a}$ which gives an astrometric centroid shift
more than a threshold amount i.e. $\delta_{T}$ is given by:
\begin{eqnarray}\label{ua}
u_{a}&=&\sqrt{\frac{T_{obs}v_{t}}{\delta_{T}D_{l}}}=25.4
\sqrt{\frac{T_{obs}(yr)}{5}}\sqrt{\frac{5\sqrt{2}\sigma_{a}}{\delta_{T}}}\times\nonumber\\&\times&\sqrt{\frac{v_{t}(km/s)}{130}}\sqrt{\frac{0.1
}{D_{l}(Kp)}},
\end{eqnarray}
where $T_{obs}$ is the lifetime of the satellite and $v_{t}$ is the
relative velocity of source with respect to the lens
\cite{DominikSahu}. Here, we assume that the astrometric centroid
shifts more than $\delta_{T}=5\sqrt{2}\sigma_{a}$ are measurable
with Gaia in which $\sigma_{a}$ is the Gaia astrometric precision.
Since Gaia can measure the astrometric displacement only along the
scan, we combine $\sigma_{a}$ with a factor of $\sqrt{2}$ to give
the Gaia two-dimensional accuracy \cite{Belokurov2002}. The amount
of the threshold impact parameter for a typical astrometric
microlensing event being observed with Gaia with the astrometric
precision $\sigma_{a}=300~\mu as$ is equal to $u_{a}\sim25$ whereas
the threshold impact parameter for the photometric observation of a
microlensing event is about one. Hence, the astrometric
cross-section which is proportional to $\propto u_{a}^{2}$ is much
larger than the photometric one.

\subsection{Astrometric optical depth}\label{optical}
According to the definition of the threshold impact parameter
$u_{a}$, during the observational time $T_{obs}$ the effective area
around each lens which yields an astrometric shift in the projected
source positions more than the threshold amount $\delta_{T}$ is
$\mathcal{A}(x,M_{l})=2 u_{a}R_{E}T_{obs}v_{t}$. This area is a
function of the lens mass $M_{l}$ and the lens distance $D_{l}=x
D_{s}$ from the observer. The astrometric optical depth is defined
as the cumulative fraction of these areas in the lens plane, for all
lens distances from the observer to the source star which is given
by \cite{DominikSahu}:
\begin{eqnarray}
\tau_{a}=D_{s}\int_{0}^{1} dx \rho_{d}(x)\int_{0}^{\infty} dM_{l}
\frac{f(M_{l})}{M_{l}} \mathcal{A}(x,M_{l}),
\end{eqnarray}
where $f(M_{l})$ is the distribution function of the lens mass and
$\rho_{d}(x)$ is the stellar density distribution in the Galactic
disk. We use the following distribution for the Galactic
disk\cite{ero09}:
\begin{eqnarray}\label{rhod}
\rho_{d}(x)=\frac{\Sigma}{2H}\exp(\frac{-(R-R_{\odot})}{h})
\exp(\frac{-|z|}{H}),
\end{eqnarray}
where $R$ is the radial distance from the Galactic center, $z$ is
the altitude with respect to the Galactic plane, $H=0.325~kpc$ is
the height scale, $h=3.5~kpc$ is the length scale of the disk and
$\Sigma$ is the column density of the disk at the Sun position. By
inserting the amount of $\mathcal{A}(x,M_{l})$, the optical depth
is:
\begin{eqnarray}
\tau_{a}=4\sqrt{\frac{G}{c^2}}D_{s}\sqrt{\frac{T_{obs}^3
v_{t}^3}{5\sqrt{2}\sigma_{a}}}\int_{0}^{1}&dx&
\rho_{d}(x)\sqrt{1-x}\nonumber\\ \int_{0}^{\infty}&
dM_{l}&\frac{f(M_{l})}{\sqrt{M_{l}}}.
\end{eqnarray}
The optical depth can be used to estimate the number of observed
events. If $N_{bg}$ background source stars brighter than $20$ G
magnitudes are observed during the Gaia lifetime $T_{obs}$, the
number of source stars with the astrometric shift, in their
projected positions, more than $\delta_{T}$ is given by:
\begin{eqnarray}\label{NEE}
N_{a}=\frac{\pi}{2}\frac{T_{obs}N_{bg}}{t_{E}u_{a}}\tau_{a}.%~\varepsilon,
\end{eqnarray}
By considering $\varepsilon$ as the efficiency for measuring the
lens mass of these observed events, the number of astrometric
microlensing events being observed with Gaia in which the lens mass
can be measured is $N_{e}=N_{a}~\varepsilon$.

%%%%%%%%%%%%%%%%%%%%%%%%%%%%%%%%%%%%%%%%%%%%%%%%%%%%%%%%%%%%%%%%%%%%%%%%%%%%%%%%%%%%%%%%%%%%%%%%%%%%%%%%%5
\section{Binary astrometric microlensing during the Gaia mission}\label{two}
The binary fractions of massive stellar populations have not
explicitly been specified yet. However, this issue is very important
for studying the gravitational waves. Because, the sources of the
gravitational waves are binary systems composed of white dwarfs,
neutron stars and black holes e.g. Riles (2013), Cutler and Thorne
(2002). Here, we want to investigate whether Gaia can specify the
binary fractions for these stellar populations through binary
astrometric microlensing. For this goal, we estimate the number of
detectable binary events which have considerable binary signals
during the Gaia era by performing Monte Carlo simulation according
to the Gaia observing strategy.

In the first step, we produce an ensemble of binary astrometric and
photometric microlensing events due to massive stellar populations
as microlenses according to their physical distributions explained
in subsection (\ref{param}). We consider binary stellar-mass black
holes, neutron stars, white dwarfs and main-sequence stars as
microlenses. Then, corresponding light curves and astrometric
trajectories are re-generated according to the Gaia observing
strategy. By considering a suitable criterion explained in
subsection (\ref{result1}), we investigate whether the binary
signatures can be inferred from the Gaia observations. Finally, by
calculating the optical depth (explained in subsection
\ref{optical}), the number of detectable binary events during the
Gaia mission are estimated. The
results are gathered in subsection (\ref{result1}).\\

\subsection{parameter space}\label{param}
The distribution functions used to generate the binary astrometric
and photometric microlensing events are explained here.

For the source stars, we make an ensemble of the source stars
according to their mass, age, metallicity and magnitude and then
choose uniformly the source star from that. We take the ages of the
stars in the range of $\log(t/yr)=[6.6, 10.2]$ with intervals of
$\Delta\log(t)=0.05~dex$. For each age range, we estimate the number
of stars according to a positive star formation rate $dN/dt>0$
\cite{Cignoni2006,Karami2010}. For each star in this age range, we
take the mass of source star, i.e. $M_{\star}$, from the Kroupa mass
function in the range of $M_{\star}\in[0.1,3]M_{\odot}$
\cite{kroupa93,kroupa01} and metallicity, i.e. $Z$, according to the
age$-$metallicity relation, and the distribution function for the
metallicity from Twarog (1980). Finally, we use the theoretical
isochrones of the colour$-$magnitude diagram (CMD) obtained using
the numerical simulation of the stellar evolution from the Padova
model \cite{Marigo08} to simulate the absolute magnitude of each
star. The isochrones are defined for stars with different masses,
colours and magnitudes but with the same metallicities and ages. %We
%select isochrones with the metallicity range of
%$Z=0.0004,~0.008,~0.004,~0.001,~0.019,~0.030$.

The coordinate of the source star toward the Galactic bulge
$(l,b,D_{s})$, is selected from $dN/d\Omega \propto\rho(l,b,D_{s})
D_{s}^{2}$ in the range of $l\in[-10,10]^{\circ}$,
$b\in[-5,5]^{\circ}$ and $D_{s}\in[0,10]~kpc$ where
$\rho(l,b,D_{s})$ is the combination of the stellar densities in the
Galactic bulge and disk \cite{ero09}. Galactic bulge has a triaxial
structure rotated along the main axes of bulge according to the
orientation angles: $\eta=78.9^\circ$ represents the angle between
the bulge major axis and the line perpendicular to the Sun-Galactic
center line, $\beta=3.5^\circ$ is the angle between the bulge plane
and the Galactic plane and $\gamma=91.3^\circ$ is the roll angle
around the major axis of bulge \cite{besancon}.

We obtain the apparent magnitude of the source star $(m_{\star})$
according to the absolute magnitude, distance modulus and
extinction. The extinction map towards the Galactic bulge and in the
Galactic distance is obtained in some references, e.g. Gonzalez et
al. (2012), Nataf et al. (2013). But the extinction map in the
closer distances is not clear. We know that the extinction can be
determined according to the column density of Hydrogen content of
the Galaxy \cite{we01}. The different fractions of Inter Stellar
Medium (ISM) in different directions toward the Galactic bulge are
due to Hydrogen content. Hence, we need to estimate these fractions
in different directions toward the Galactic bulge. We do this by
comparing (in the each direction at the Galactic distance) (i) the
extinction map provided by Gonzalez et al. (2012) and (ii) the
extinction amount calculated according to its relation with the
Hydrogen column density by assuming all ISMs are made from the
Hydrogen content. We use ISM density law introduced by Robin et al.
(2003). Also we obtain the V-band extinction from its value in near
infrared bands according to the relations introduced by Cardelli et
al. (1989). Indeed, we assume the fraction of ISM composed from the
Hydrogen content in the Galactic scale depends only on the direction
but not on the distance. The source stars fainter than $20$ G
magnitudes are not considered.

For generating the light curve of microlensing events with the
finite source size effect, we need the radius of source star. For
the main-sequence stars, the relation between the mass and radius is
given by $R_{\star}= M_{\star}^{0.8}$ where all parameters
normalized to the sun's value. We do not consider giant source
stars. The Gaia astrometric and photometric precisions are pointed
out according to the apparent magnitude of source \cite{Varadi2009}.
We consider the events in which $\theta_{E}>5\sqrt{2}\sigma_{a}$.

For indicating the trajectory of the source with respect to x-axis,
we identify this path with an impact parameter and orientation
defined by an angle between the trajectory of source star and x-axis
i.e. $\xi$. We let this angle change in the range of $[0,2 \pi]$.
The impact parameter is taken uniformly in the range of
$u_0\in[-u_{a},u_{a}]$ and the corresponding time for $u_0$ is
chosen uniformly in the range of $t_{0}\in[0,5]$ years which is the
lifetime of Gaia.

To simulate the position of the lens in the Galactic disk, by
considering the same Galactic latitude $b$ and longitude $l$ as the
source, we take its distance with respect to the observer from the
probability function of astrometric microlensing detection
$d\Gamma/dx\propto \rho_{d}(x)(1-x)^{2}$, where $x=D_{l}/D_{s}$
\cite{DominikSahu}. %changes in the range of $x\in [0,1]$ .

The mass of the stellar-mass black holes as microlenses is taken
from the distribution function introduced by Farr et al. (2011). We
take the mass of the neutron stars from a Gaussian function, with an
average and dispersion of $M_{0}=1.28~M_{\odot}$ and
$\sigma=0.28~M_{\odot}$ in the range of $[1,2.5]~M_{\odot}$
\cite{Ozel2012}. The mass of white dwarfs is taken from a
combination of two Gaussian functions with the averages and
dispersions of $M_{0}=0.59~M_{\odot}$, $\sigma=0.047~M_{\odot}$ and
$M_{0}=0.711~M_{\odot}$, $\sigma=0.109~M_{\odot}$ for two different
samples of white dwarfs in the range of $[0.3,1.4]~M_{\odot}$
\cite{Kepler2007}. We take the mass of main-sequence stars as
microlenses from the Kroupa mass function in the range of
$[0.1,2]~M_{\odot}$.

The velocities of the lens and the source star are taken from the
combination of the global and dispersion velocities of the Galactic
disk and bulge \cite{ero09,bin}. %The relative velocities of the
%source-lens is determined by projecting the velocity of the source
%star into the lens plane.

%%%%%%%%%%%%%%%%%%%%%%%%%%%%%%%%%%%%%%%%%%%%%%
\begin{figure}
\includegraphics[angle=270,width=8.cm,clip=]{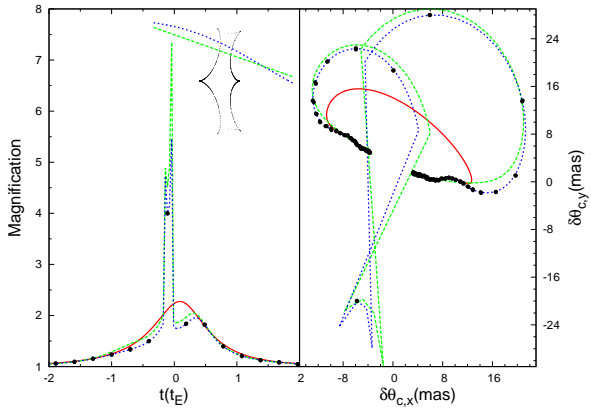}
\caption{A typical binary microlensing event with binary
stellar-mass black holes. The light curves (left panel), astrometric
trajectories (right panel) and the source trajectories with respect
to the caustic curve (the inset in the left-hand panel) without and
with considering the parallax effect are shown with green dashed and
blue dotted lines respectively. The point-mass lens models
(considering the parallax effect) are shown by red solid lines. Data
points taken by Gaia are shown with black points. The relevant
parameters of the source and lens are $M_{\star}=1.1~M_{\odot}$,
$m_{\star}=15.8~mag$, $D_{s}=2.0~kpc$, $\xi=-20^{\circ}$,
$v_{t}=120~km s^{-1}$, $u_{0}=0.4$ and $M_{l}=10~M_{\odot}$,
$q=0.6$, $D_{l}=0.4~kpc$, $d=0.9~R_{E}$ respectively.} \label{fig1}
\end{figure}
%%%%%%%%%%%%%%%%%%%%%%%%%%%%%%%%%%%%%%%%%%%%%%

To simulate  the parallax effect, we first convert the Galactic
coordinate of the source to the ecliptic coordinate using LAMBDA
site\footnote{http://lambda.gsfc.nasa.gov/toolbox/tb-coordconv.cfm}.
Then, the Earth's position vector with respect to the Sun is
calculated (e.g. the Appendix section from An et al. (2002)). The
time when the Earth is in the perihelion is $t_{e}=-18$days with
respect to the time when Gaia entered its operational orbit i.e.
start time of observation.

Here, we review the Gaia scanning law. Gaia has two astrometric
fields of view separated by $106.5^{\circ}$. Its spin axis makes an
angle of $45^{\circ}$ with respect to the direction towards the Sun.
The satellite spins around this axis with the angular rate some of
$60~as~s^{-1}$. Hence, observations occur in pair whose time
interval is $106.5$ minutes. The spin axis in turn has a precession
motion around a circle. Finally, the spacecraft will move in a
Lissajous orbit around the second Lagrange point of the Sun-Earth
system. The combination of these motions gives the Gaia scanning law
\footnote{http://sci.esa.int/gaia/40846-galactic-structure}. Each
object will be observed $72$ times averagely with the cadence some
of $25$ days. This satellite measures the positions of objects just
set along the scan direction. For simulating the observational
astrometric trajectories and photometric light curves, we adopt the
Gaussian distribution for the cadence between two pairs of data
taken by Gaia with the average and dispersion of $25$ and $5$ days.
Each pair occurs in $106.5$ minutes. In each observation we consider
a random scanning direction chosen uniformly in the range of
$[0,2\pi]$.

Here, we explain distribution functions used for binary systems. For
all binary systems we take the mass ratio from the distribution
function introduced by Duquennoy \& Mayor (1991). We consider binary
stellar-mass black holes (BBH), binary neutron stars (BNS), binary
white dwarfs (BWD) and binary main-sequence stars (BMS) as
microlenses. (Note that, BNS and BWD are not necessarily double
neutron stars and double white dwarfs. Their primary lenses are a
neutron star and a white dwarf respectively.) We choose the
semi-major axis $s$ of the binary orbit from the \"{O}pik's law
where the distribution function for the primary-secondary distance
is proportional to $\rho(s)=dN/ds\propto s^{-1}$ \cite{Opik} in the
range of $s\in[0.6,30]~A.U.$. We use the generalized version of the
adaptive contouring algorithm \cite{Dominik2007} for simulating
astrometric trajectories in binary microlensing events.\\

We let the microlenses rotate around their center of mass. Indeed,
the perturbations developed with time such as lens orbital motion
can make considerable deviations in astrometric trajectories which
helps to discover the second microlens \cite{Sajadian}. The
parameters of lenses orbit are taken as follows: the time of
arriving at the perihelion point of orbit $t_{p}$ is chosen
uniformly in the range of $t_{p}\in [t_{0}-P,t_{0}+P]$ where $P$ is
the orbital period of lenses motion. The projection angles to
specify the orientation of orbit of lenses with respect to the sky
plane, i.e. $\delta$ and $\vartheta$, are taken uniformly in the
range of $[-\frac{\pi}{2},\frac{\pi}{2}]$. We take the eccentricity
of the lenses' orbit uniformly in the range of $\emph{e}\in[0,1]$.
%%%%%%%%%%%%%%%%%%%%%%%% table 1 %%%%%%%%%%%%%%%%%%%%%%%%%%
\begin{deluxetable*}{cccccccccc}
\tablecolumns{10} \centering \tablewidth{\textwidth} \tabletypesize
\footnotesize \tablecaption{The results of the simulation of the
binary astrometric microlensing events with GAIA.} \tablehead{
\colhead{} & \colhead{$\overline{t_{E}} (days)$} &
\colhead{$\overline{t_{ae}} (days)$} & \colhead{$\overline{u_{a}}$}
& \colhead{$\varepsilon_{b} (\%)$}&
\colhead{$\Sigma~(M_{\odot}~pc^{-2})$}&
\colhead{$\overline{\tau_{a}}\times 10^{-7}$}&
\colhead{$N_{b}(f=0.1)$}& \colhead{$N_{b}(f=0.3)$}&
\colhead{$N_{b}(f=0.5)$}}

\startdata
$BBH$& $90.2$ & $568.6$ & $18.3$ & $13.5$ & $2.09$ & $8.0$ & $5.6$ & $16.9$ & $28.1$ \\
\\
$BNS$& $27.1$ & $109.5$ & $32.8$ & $16.5$ & $0.91$ & $8.6$ & $11.4$ & $34.2$ & $57.1$ \\
\\
$BWD$& $17.9$ & $66.0$ & $34.1$ & $15.6$ & $2.41$ & $34.8$ & $76.8$ & $230.3$ & $383.8$ \\
\\
$BMS$& $11.5$ & $42.5$ & $43.3$ & $17.6$ & $21.3$ & $434.9$ & $1316.1$ & $3948.4$ & $6580.6$ \\
\enddata
\tablecomments{The average values of the Einstein crossing time
$\overline{t_{E}} (days)$, the duration of the astrometric event
$\overline{t_{ae}}~(days)$ threshold impact parameter
$\overline{u_{a}}$, efficiency of detecting binary signatures of
microlenses $\varepsilon_{b} (\%)$, the column density
$\Sigma~(M_{\odot}~pc^{-2})$, astrometrical optical depth
$\overline{\tau_{a}}$ and number of binary astrometric microlensing
events detected with Gaia due to different binary systems: binary
stellar-mass black holes (BBH), binary neutron stars (BNS), binary
white dwarfs (BWD) and binary main-sequence stars (BMS) considering
different binary fractions $f$.} \label{tab2}
\end{deluxetable*}
%%%%%%%%%%%%%%%%%%%%%%%%%%%%%%%%%%%%%%%%%%%%%%%%%%%%%%%%%%%%%

\subsection{results}\label{result1}
To determine the detectability of binary signals we calculate
$\chi^{2}$. Two models are fitted to the astrometric and photometric
data: (a) the known rotating binary microlensing model with the
parameters used to simulate data, i.e. $\chi^{2}_{b}$ and (b) the
simple Paczy\'nski microlensing events with the same Paczy\'nski
parameters, i.e. $\chi^{2}_{P}$. Our criterion for detectability of
the binary signal is $\Delta
\chi^{2}=\chi^{2}_{P}-\chi^{2}_{b}>\Delta\chi^{2}_{th}$.
$\chi^{2}_{j}$ (for j=b,P) from fitting the astrometric trajectory
along the scan $\theta_{s}$ and the magnification factor $A$ is
given by:
\begin{eqnarray}
\chi^{2}_{j}=\Sigma_{i=1}^{N}[\frac{(A_{i}-A_{j}(t_{i}))^{2}}{\sigma_{A}^{2}}+
\frac{(\theta_{s,i}-\theta_{s,j}(t_{i}))^{2}}{\sigma_{a}^{2}}]
\end{eqnarray}
where $N$ is the number of data points and $\sigma_{A}$ is the Gaia
photometric precision. We assume that from fitting process and
searching all parameter space the best-fitted solution is the known
binary solution. We consider $\Delta\chi^{2}_{th}=200$.

In Figure (\ref{fig1}) we show a typical simulated binary
microlensing event with binary stellar-mass black holes being
observed with Gaia. The light curves (left panel), astrometric
trajectories (right panel) and the source trajectories with respect
to the caustic curve (the inset in the left-hand panel) without and
with considering the parallax effect are shown with green dashed and
blue dotted lines respectively. The point-mass lens models are shown
by red solid lines. Data points taken by Gaia are shown with black
points. The relevant parameters of the source and lens are
$M_{\star}=1.1~M_{\odot}$, $m_{\star}=15.8~mag$, $D_{s}=2.0~kpc$,
$\xi=-20^{\circ}$, $v_{t}=120~km s^{-1}$, $u_{0}=0.4$ and
$M_{l}=10~M_{\odot}$, $q=0.6$, $D_{l}=0.4~kpc$, $d=0.9~R_{E}$
respectively. In this event, the binary signal is detectable with
Gaia. Although, in the photometric light curve there is one data
point with significant deviation with respect to the simple
microlensing light curves but in the astrometric trajectory there
are several deviated data points. Hence, although the Gaia cadence
is too sparse so that some of caustic-crossing features may be
missed in the photometric light curves but the astrometric
measurements can much likely resolve the binary signatures from the
second microlenses. Indeed, the astrometric cross section is much
larger than the photometric one.

In Table (\ref{tab2}) we provide the average values of the Einstein
crossing time $\overline{t_{E}}$, the duration of the astrometric
event $\overline{t_{ae}}$, threshold impact parameter
$\overline{u_{a}}$, efficiency for detecting binary signals
$\varepsilon_{b}$, i.e. the fraction of the simulated binary events
in which the binary signals are detectable with
$\Delta\chi^2>\Delta\chi^{2}_{th}$ given in per cent, the column
densities for different stellar populations $\Sigma$ (taken from
Flynn et al. (2006) and Chabrier (2001)) and the astrometrical
optical depth $\overline{\tau_{a}}$ from the second to sixth column
due to different binary systems. Note that, we calculate the optical
depth for different directions toward the Galactic bulge, i.e. $l$
in the range of $[-10,10]^{\circ}$ and $b$ in the range of
$[-5,5]^{\circ}$ and average over all of them.

The expected number of binary microlensing events being observed
with Gaia with detectable binary signals, i.e. $N_{b}$, is given by:
\begin{eqnarray}
N_{b}=N_{a}~f~\varepsilon_{b}, %\frac{\pi}{2}\frac{T_{obs}N_{bg}f}{\overline{t_{E}}\overline{u_{a}}}\overline{\tau_{a}}
\end{eqnarray}
where $f$ is the fraction of binary objects in different stellar
populations. To calculate $N_{a}$ (equation \ref{NEE}) we insert the
average values of $\overline{t_{E}}$, $\overline{u_{a}}$ for each
stellar population. Also, we set the number of background source
stars towards the Galactic bulge $N_{bg}=3.0\times 10^{8}$
\cite{OGLE}. Noting that the number of background stars brighter
than $20$-G magnitudes for whole sky is on the order of billion
objects. The amounts of $N_{b}$ for three values of $f=0.1,0.3$ and
$0.5$ are reported in Table (\ref{tab2}).

According to this table: (i) The Gaia efficiency for detecting
different binary systems as microlenses is almost $10-20$ per cent
which is acceptable according to the sparse cadence of Gaia. It
means that Gaia detects the signal from the second lens with
probability $\sim 10-20$ per cent.

(ii) The Gaia efficiency decreases with increasing the lens mass.
Because, by increasing the lens mass, the normalized distance
between two lens decreases. Hence, binary black holes usually make
close binaries for which the probability of caustic crossing is
smaller than that in the intermediate binaries due to the less
massive microlenses.

(iii) Gaia can potentially determine the binary fraction of massive
stellar populations. Because, the numbers of estimated events for
different binary fractions are larger than one. The number of
estimated events for binary black holes is small which means that
the uncertainty for specifying their binary fraction is high,
specially if the real binary fraction would be less than $0.1$.

(iv) The duration of the astrometric events is given by
\cite{DominikSahu,Belokurov2002}:
$\overline{t_{ae}}=t_{E}~\theta_{E}/(5\sqrt{2}\sigma_{a})$ which is
proportional to the Einstein crossing time. Indeed, the ratio of
$\theta_{E}/5\sqrt{2}\sigma_{a}$ is almost constant for different
stellar populations and larger than one. Hence, the more massive
microlenses with longer Einstein crossing time have the longer
astrometric durations. The longest duration of the astrometric
events is that due to binary stellar-mass black holes and some of
$1.56~yrs$. This time is one third of the Gaia lifetime. The more
massive microlenses have longer astrometric duration, comparable
with the Gaia lifetime which are not suitable to be observed with
Gaia.

%%%%%%%%%%%%%%%%%%%%%%%% table 2 %%%%%%%%%%%%%%%%%%%%%%%%%%
\begin{deluxetable*}{cccccccccc}
\tablecolumns{10} \centering \tablewidth{\textwidth} \tabletypesize
\footnotesize \tablecaption{The results of the simulation of
astrometric microlensing events due to massive stellar populations
with GAIA.} \tablehead{ \colhead{} &
\colhead{$\overline{M_{l}}~(M_{\odot})$} &
\colhead{$\overline{t_{E}}~(days)$} &
\colhead{$\overline{t_{ae}}~(days)$} & \colhead{$\overline{u_{a}}$}&
\colhead{$\overline{v_{t}}~(km~s^{-1})$}&
\colhead{$\overline{D_{s}}~(kpc)$}&
\colhead{$\overline{\sigma_{a}}~(\mu~as)$}&
\colhead{$\varepsilon~(\%)$}& \colhead{$N_{e}$} } \startdata
\\
$(BH)$& $10.3$ & $74.1$ & $460.2$ & $20.1$ & $179.8$ & $4.3$ & $478.8$ & $9.8$ & $44.5$ \\
\\
$(NS)$& $1.4$ &  $20.5$ & $86.7$ &  $29.0$ & $182.5$ & $3.5$ &$324.2$& $2.9$ & $34.2$ \\
\\
$(WD)$& $0.6$ & $12.3$ & $46.6$ & $35.0$ & $182.7$ & $3.1$ &$268.9$ & $1.2$ & $76.3$ \\
\\
$(MS)$& $0.4$ & $9.1$ & $33.3$ & $41.0$ & $182.8$ & $2.7$ & $232.1$ & $0.8$ & $785.8$ \\
\enddata
\tablecomments{The average values of the lens mass
$\overline{M_{l}}~(M_{\odot})$, Einstein crossing time
$\overline{t_{E}}~(days)$, duration of the astrometric event
$\overline{t_{ae}}~(days)$, threshold impact parameter
$\overline{u_{a}}$, lens-source transverse velocity
$\overline{v_{t}}~(km~s^{-1})$, source distance from the observer
$\overline{D_{s}}~(kpc)$, astrometric precision of Gaia $
\overline{\sigma_{a}}~(\mu as)$, detection efficiency for
high-quality events $\varepsilon~(\%)$ and finally the number of
high-quality events being observed during the Gaia era for four
different stellar populations: stellar-mass black holes (BH),
neutron stars (NS), white dwarfs (WD) and main-sequence stars (MS).}
\label{tab1}
\end{deluxetable*}
%%%%%%%%%%%%%%%%%%%%%%%%%%%%%%%%%%%%%%%%%%%%%%%%%%%%%%%%%%%%%
%%%%%%%%%%%%%%%%%%%%%%%%%%%%%%%%%%%%%%%%%%%%%%%%%%%%%%%%%%%%%
\section{Measuring the mass of massive stellar populations}\label{three}
Here we investigate how Gaia is efficient to measure the mass of the
massive stellar populations through astrometric microlensing of
single lenses and specify their mass distributions. In this regard,
we quantitatively estimate the Gaia efficiency and the number of
high-quality astrometric microlensing events by performing a Monte
Carlo simulation.

We consider different stellar populations, i.e. stellar-mass black
holes, neutron stars, white dwarfs and main-sequence stars as
microlenses and simulate astrometric microlensing events due to
these microlenses according to their distribution functions
explained in subsection (\ref{param}). Having re-generated the
astrometric and photometric curves with synthetic data points
according to the Gaia observing strategy, we should indicate whether
the lens mass can be inferred from data. In that case, we evaluate
the uncertainty in the lens mass i.e. $\sigma_{M}$, in the same way
as Belokurov \& Evans (2002), and accept the events with the
relative error i.e. $\sigma_{M}/M_{l}$ less than $0.5$ as
high-quality
events. \\

\subsection{results}
The results of this Monte Carlo simulation are summarized in Table
(\ref{tab1}) which lists the average values of the lens mass
$\overline{M_{l}}~(M_{\odot})$, Einstein crossing time
$\overline{t_{E}}~(days)$, duration of the astrometric event
$\overline{t_{ae}}~(days)$, threshold impact parameter
$\overline{u_{a}}$, lens-source transverse velocity
$\overline{v_{t}}~(km~s^{-1})$, source distance from the observer
$\overline{D_{s}}~(kpc)$, astrometric precision of Gaia
$\overline{\sigma_{a}}~(\mu~as)$, efficiency for detecting
high-quality events $\varepsilon~(\%)$, i.e. the number of the
high-quality events divided by the total number of simulated events
given in per cent and finally the number of high-quality events
$N_{e}$ (explained in subsection \ref{optical}) for four different
stellar populations.%: stellar-mass black holes (BH), isolated neutron
%stars (NS), white dwarfs (WD) and main-sequence stars (MS).

Interpreting this table: (i) The more massive lenses, the larger
efficiencies of measuring the lens mass. Indeed, the Einstein
crossing time in astrometric microlensing events due to massive
lenses is long. As a result, the number of data over the astrometric
trajectories and photometric light curves increases. On the other
hand, their angular Einstein radius and astrometric signal are high.

(ii) The Gaia astrometric precision for astrometric microlensing due
to more massive microlenses is larger than that due to less massive
ones. Indeed, we consider two restrictions for simulating
astrometric microlensing events: (a) the source star should be
brighter than 20 G magnitude and (b) the angular Einstein radius
should be larger than $5 \sqrt{2} \sigma_{a}$. Hence, the average
value of $\sigma_{a}$, reported in seventh column, represents the
minimum amount of the angular Einstein radius which in turn is an
increasing function with respect to the lens mass. As a result, the
threshold impact parameter which is proportional to
$1/\sqrt{\delta_{T}}$ decreases with enhancing the lens mass.

(iii) The number of high-quality events due to the massive stellar
populations are larger than one. Hence, we can obtain some
information about their masses and mass distribution. A significant
point is that this method can even give the mass of isolated
objects. For the main-sequence stars as microlenses, we conclude
that some of $786$ high-quality events \emph{toward the Galactic
bulge} will probably be confirmed during the Gaia era. However,
Belokurov and Evans (2002) estimated that number of high-quality
events due to main-sequence stars \emph{for whole sky} is $\sim
2500$. Indeed, we consider the number of background stars brighter
than $20$ G magnitudes towards the Galactic bulge as $300$ million
objects which is so fewer than that was considered by them, i.e. one
billion objects. Meantime, there are some differences between the
distribution functions used in our simulation and those used in the
referred paper. However, they estimated just high-quality events due
to main-sequence stars.

Note that the mass ranges of these four stellar populations as
microlenses overlap in edges. But, we can discern the type of
microlenses with the masses in the common ranges (belong to two
populations) through their color and magnitude. Because, these
microlenses most likely locate at the distance less than $1$ kilo
parsec from the observer.
%%%%%%%%%%%%%%%%%%%%%%%%%%%%%%%%%%%%%%%%%%%%%%%5
\begin{figure*}
\subfigure[] {
\includegraphics[angle=0,width=8.cm,clip=]{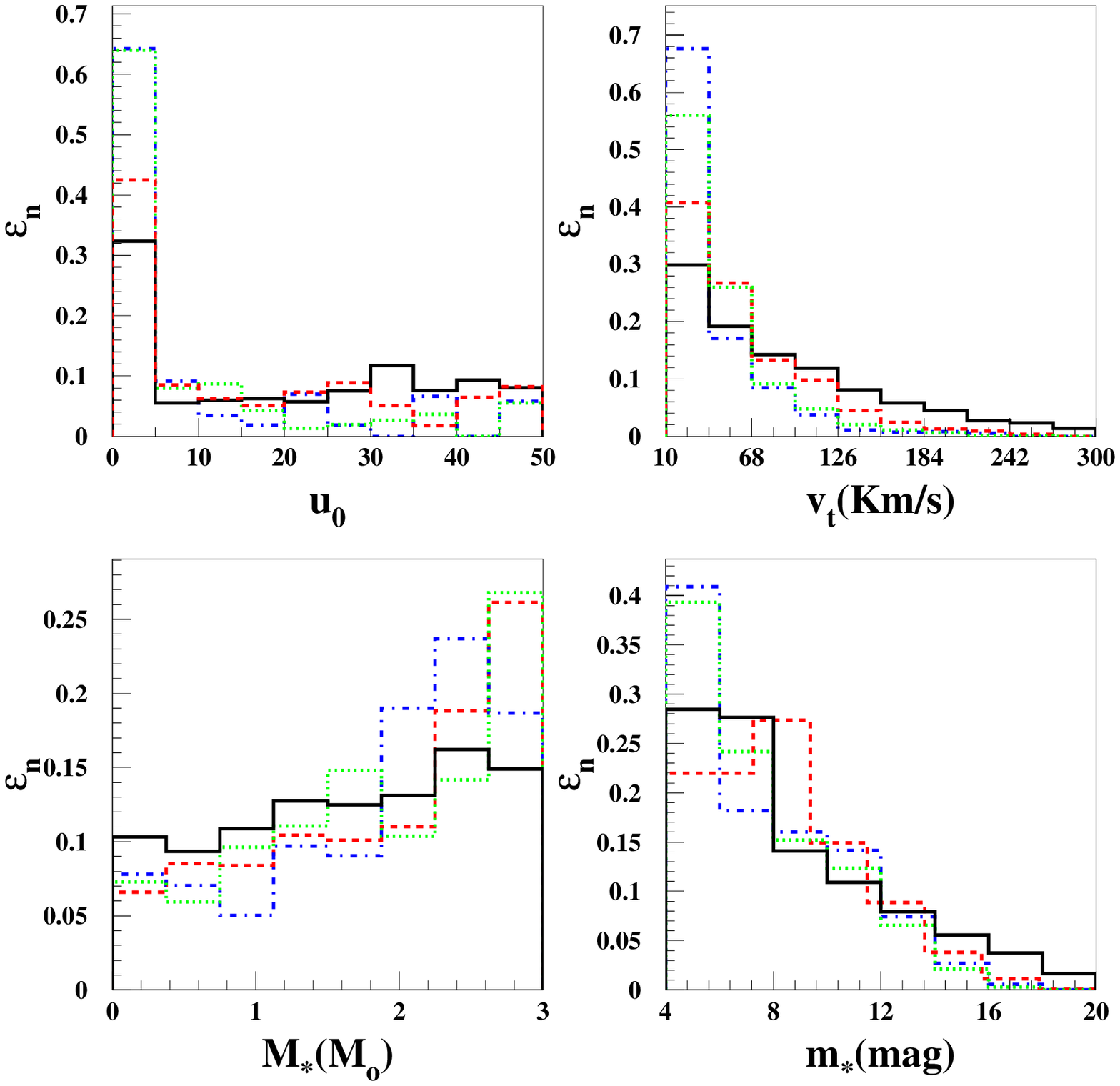}
} \subfigure[] {
\includegraphics[angle=0,width=8.cm,clip=]{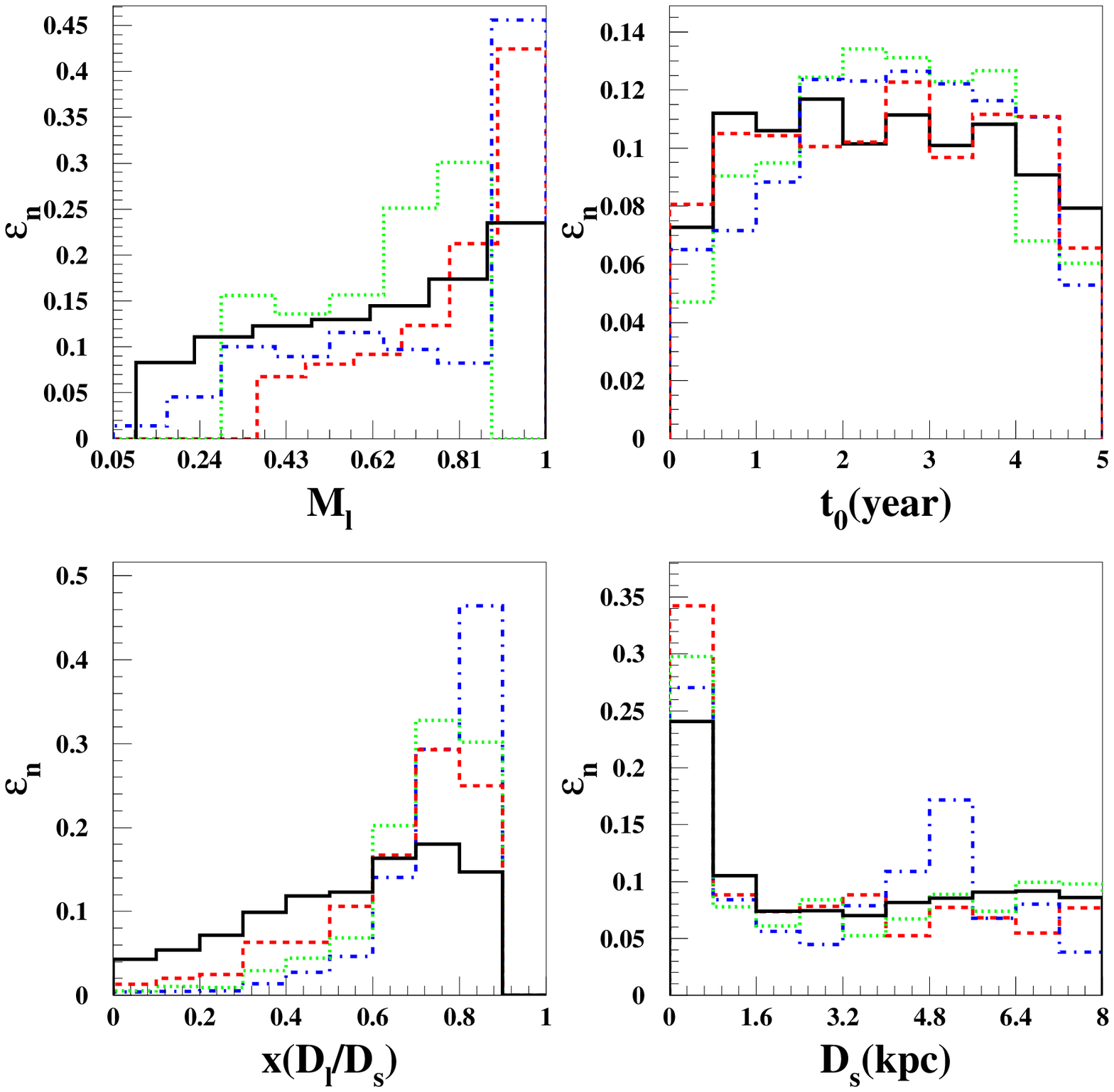} }
\caption{The normalized efficiency curves, $\varepsilon_{n}$, of
Gaia for measuring the lens mass in astrometric microlensing due to
stellar-mass black holes (black solid line), neutron stars (red
dashed line), white dwarfs (green dotted line) and main-sequence
stars (blue dot-dashed line) for some parameters of source star and
lens. Noting that all efficiency curves are normalized to one which
means the area under each curve is one, so-called as normalized
efficiency curves. Also, the lens mass for each stellar population
is divided to its maximum amount so that the range of the normalized
lens mass for all populations is in the range of $[0,1]$.}
\label{fig5}
\end{figure*}
%%%%%%%%%%%%%%%%%%%%%%%%%%%%%%%%%%%%%%%%%%%%%%

In order to study the dependance of efficiency on the parameters of
the model, in Figure (\ref{fig5}) we plot the normalized
efficiencies, i.e. $\varepsilon_{n}$, for detecting high-quality
events due to stellar-mass black holes (black solid line), neutron
stars (red dashed line), white dwarfs (green dotted line) and
main-sequence stars (blue dot-dashed line) in terms of the relevant
parameters of the lens and source. To compare the efficiency curves
with each other, we normalize all efficiency curves to one which
means the area under each curve is one, hence we name them as
\emph{normalized efficiency curves}. Since the area under each
efficiency curve is proportional to the related value reported in
the ninth column of Table (\ref{tab1}) i.e. $\varepsilon$, the
normalization is done by dividing Y values to $\varepsilon$.
Generally, the behaviors of efficiency for different stellar
populations are similar, although some differences exist in the
slope of curves, peak position, etc. The normalized efficiency
function in terms of the lens and source parameters is given as
follows:

(i) The first one is the impact parameter, $u_{0}$. The efficiency
of detecting high-quality events is high for astrometric
microlensing events with $u_{0}<1$ for which the photometric
microlensing can be observed as well as the astrometric one.

(ii) The second parameter is the relative velocity of source with
respect to lens, $v_{t}$. The efficiency increases with decreasing
the relative velocity. Indeed, the Einstein crossing time rises with
decreasing the relative velocity. Since the Gaia cadence is almost
constant for all events towards the Galactic bulge, so the number of
observational data for long-duration microlensing events (over the
astrometric trajectories and photometric light curves) is high. As a
result, the errors in parameters for these events are smaller than
those due to short-duration ones.

(iii) and (iv) Two next parameters: $M_{\star}$ the mass and
$m_{\star}$ apparent magnitude of source star. The Gaia astrometric
precision depends strongly on the source magnitude and increases
with decreasing it. Hence, the Gaia efficiency for measuring the
lens mass in astrometric microlensing due to brighter and heavier
sources is higher than that due to fainter and less massive ones.

(v) The next parameter is the normalized lens mass, $M_{l}$. Since
the mass range for all stellar population is not similar, we
normalize the lens mass for each stellar population to its maximum
amount so that the range of the normalized lens mass for all
populations is in the range of $[0,1]$. For the massive lenses the
Einstein crossing time is long. Consequently, the number of data
over the astrometric trajectory and photometric light curve
increases. On the other hand, their angular Einstein radius and
astrometric signal are high. Therefore, the Gaia efficiency for
measuring the mass of more massive lenses is higher than that of
less massive lenses.

(vi) The next parameter is the corresponding time to the closest
approach, $t_{0}$. We uniformly choose $t_{0}$ in the range of
$[0,5]T$. But the events with $t_{0}\sim 2.5 T$ has the maximum
efficiency. Because, for these events the number of observational
data over the astrometric trajectory and light curve is maximum.

(vii) and (viii) Two last parameters: $x=D_{l}/D_{s}$ the relative
distance of the lens to the source star from the observer and
$D_{s}$ the source distance from observer. Indeed there are two
kinds of suitable sources for the Gaia detection: (a) nearby sources
with $D_{s}<1~kpc$ and (b) the bright sources often located at the
Galactic disk. Correspondingly, there are two peaks in the
efficiency diagram versus $D_{s}$. These two classes have the
different $x$s. We know that the Einstein crossing time is
proportional to $t_{E}\propto \sqrt{D_{s}x(1-x)}$, but the relative
parallax and the angular Einstein radius depend on these parameters
as $\theta_{E}\propto
\sqrt{\pi_{rel}}\propto\sqrt{\frac{1-x}{D_{s}x}}$. Although, the
efficiency enhances with increasing each of them, but the first is
an increasing function versus $x$ and $D_{s}$ and the others are
decreasing functions. Therefore, the efficiency is maximized for the
first class when $x\sim[0.6,0.7]$ and for the second class when
$x\ll1$.

According to the dependance of efficiency on the parameters, we
expect that most high-quality events have long Einstein crossing
times or small impact parameters or their source is a bright star.
Hence, most sources are so close to the observer. \\
%%%%%%%%%%%%%%%%%%%%%%%%%%%%%%%%%%%%%%%%%%%%%%
\begin{figure}
\includegraphics[angle=270,width=8.cm,clip=]{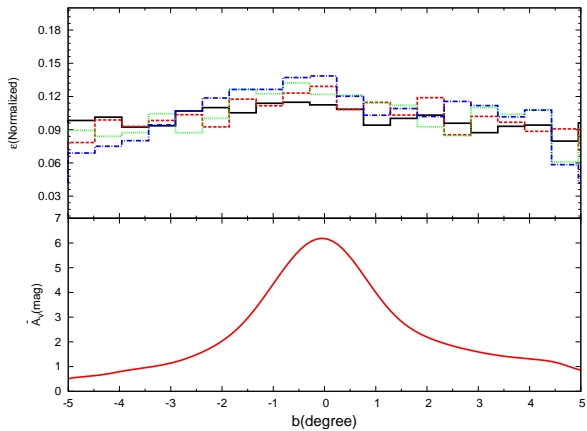}
\caption{The normalized efficiency of Gaia for measuring the lens
mass in astrometric microlensing due to different stellar
populations (top panel) and the average value of the extinction in
V-band, $\bar{A}_{v}$, (red solid line) versus the Galactic latitude
$b$. Note that the efficiency curve for each population normalized
to the related cumulative value reported in the ninth column of
Table \ref{tab1}.} \label{fign}
\end{figure}
%%%%%%%%%%%%%%%%%%%%%%%%%%%%%%%%%%%%%%%%%%%%%%

We also investigate the dependence of the Gaia (normalized)
efficiency on the Galactic latitude, shown in the top panel of
Figure (\ref{fign}). The normalized efficiency decreases very slowly
by increasing the distance from the Galactic plane. Because, there
are two effects act reversely: (i) The extinction toward the
Galactic bulge strongly increases by decreasing the distance from
the Galactic plane as shown in the bottom panel of Figure
(\ref{fign}) and (ii) the Galactic disk density, equation
(\ref{rhod}), decreases by increasing distance from the Galactic
plane. The first effect decreases the probability of finding a
source star brighter than 20-G magnitude toward the Galactic center
while the abundance of stars is maximum in this direction.
Consequently, the efficiency is almost constant over the Galactic
bulge. To determine the extinction, we average over the amounts of
the extinction due to the Galactic longitudes in the range of
$[-10,10]^{\circ}$ for each value of $b$. %This figure was provided
%by Belokurov and Evans (2002) but for whole range of $b$. Here, We
%improved its resolution toward the Galactic bulge.

%%%%%%%%%%%%%%%%%%%%%%%%%%%%%%%%%%%%%%%%%%%%%%%%%%%%%%%%%%%%%%%%%%%%%%%%%%%%%%%%%%%%%%%%%%%%%%%%%%
\section{conclusions}\label{four}
One of astronomical phenomena being observed with Gaia is
astrometric microlensing events. In this work, we investigated
whether Gaia can specify the binary fractions and mass distributions
of massive stellar populations through astrometric microlensing.

We simulated binary astrometric microlensing events being observed
with Gaia due to binary stellar-mass black holes, binary neutron
stars, binary white dwarfs and binary main-sequence stars to
evaluate the efficiency for detecting the binary signatures. We
considered $\Delta \chi^{2}>200$ as the detectability criterion and
concluded that Gaia detects the signal from the second lens with
probability $\sim10-20$ per cent. Then, We estimated the number of
binary astrometric microlensing due to the mentioned populations
with detectable binary signals in the Gaia era as $6$, $11$, $77$
and $1316$ respectively ($f=0.1$). Hence, Gaia can potentially
determine the binary fraction of massive stellar populations. Since
the predicted number of binary astrometric microlensing events of
stellar-mass black holes is small, so the Gaia accuracy to specify
their binary fraction is low. The significant points is that binary
systems composed of white dwarfs, neutron stars or black holes are
sources of detectable gravitational waves. Therefore, indicating the
binary fraction of these massive stellar populations helps to assess
the amount of the gravitational waves.

Also, we investigated the Gaia efficiency for measuring the mass of
massive stellar populations in the Galactic disk, i.e. stellar-mass
black holes, neutron stars and white dwarfs in addition to
main-sequence stars as microlenses in astrometric microlensing
events of single lenses. The Gaia efficiency for measuring the mass
of microlenses enhances with increasing the lens mass. By performing
Monte Carlo simulation, we concluded that the Gaia efficiencies for
detecting high-quality events due to these stellar populations are
$9.8$, $2.9$, $1.2$ and $0.8$ per cent respectively. We estimated
the number of high-quality events for these populations as $45$,
$34$, $76$ and $786$ respectively. Hence, Gaia can obtain some
information about their masses and mass distribution of these
massive stellar populations.

\begin{acknowledgments}
I am grateful to Sohrab Rahvar for helpful discussions and comments
and Scott Gaudi for careful reading and commenting on the
manuscript. I thank anonymous referee for useful comments and
suggestions which certainly improved the manuscript. Also, I
acknowledge the use of the Legacy Archive for Microwave Background
Data Analysis (LAMBDA), part of the High Energy Astrophysics Science
Archive Center (HEASARC).
\end{acknowledgments}

\end{document}